\documentclass{article}
\usepackage{arxiv}

\usepackage[utf8]{inputenc} 
\usepackage[T1]{fontenc}    
\usepackage[hidelinks]{hyperref}
\usepackage{booktabs}
\usepackage{amsfonts}
\usepackage{microtype}
\usepackage{graphicx}
\usepackage{url, doi}
\usepackage[yyyymmdd]{datetime}
\usepackage{kpfonts}
\usepackage{siunitx}

\title{A micro-CT of a human skull}

\author{
    \hspace{2mm}Thomas Kirchner\hspace{0.1mm} \href{https://orcid.org/0000-0002-3819-1987}{\includegraphics[scale=0.06]{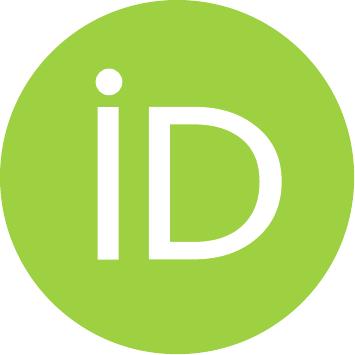}}\\
	Institut für Physik \\
	Martin-Luther-Universität Halle-Wittenberg, Halle (Saale), Germany \\
	thomas.kirchner@physik.uni-halle.de
}

\renewcommand{\headeright}{Data descriptor}
\renewcommand{\undertitle}{Data descriptor}
\renewcommand{\shorttitle}{A micro-CT of a human skull}
\date{\tomorrow}
\newdate{date}{22}{02}{2022}
\date{\displaydate{date}}

\hypersetup{
pdftitle={A micro-CT of a human skull},
pdfsubject={physics.med-ph},
pdfauthor={Thomas Kirchner},
pdfkeywords={micro-CT, X-ray microtomography, ex vivo, human, Homo sapiens, skull},
}
\begin{document}
\maketitle
\begin{abstract}
This is a data descriptor for an X-ray microtomography of a human skull. The data is available at \textbf{\doi{10.5281/zenodo.6108435}}, open data and adheres to the FAIR \cite{FAIR} principles.
\end{abstract}


\section{Data set}
The data set includes the X-ray projection images (the measured raw data), a 3D reconstruction in approximate Hounsfield units (HU) and a semi-manual bone segmentation.

\begin{figure}[hb]
    \centering
    \includegraphics[width=0.7\textwidth]{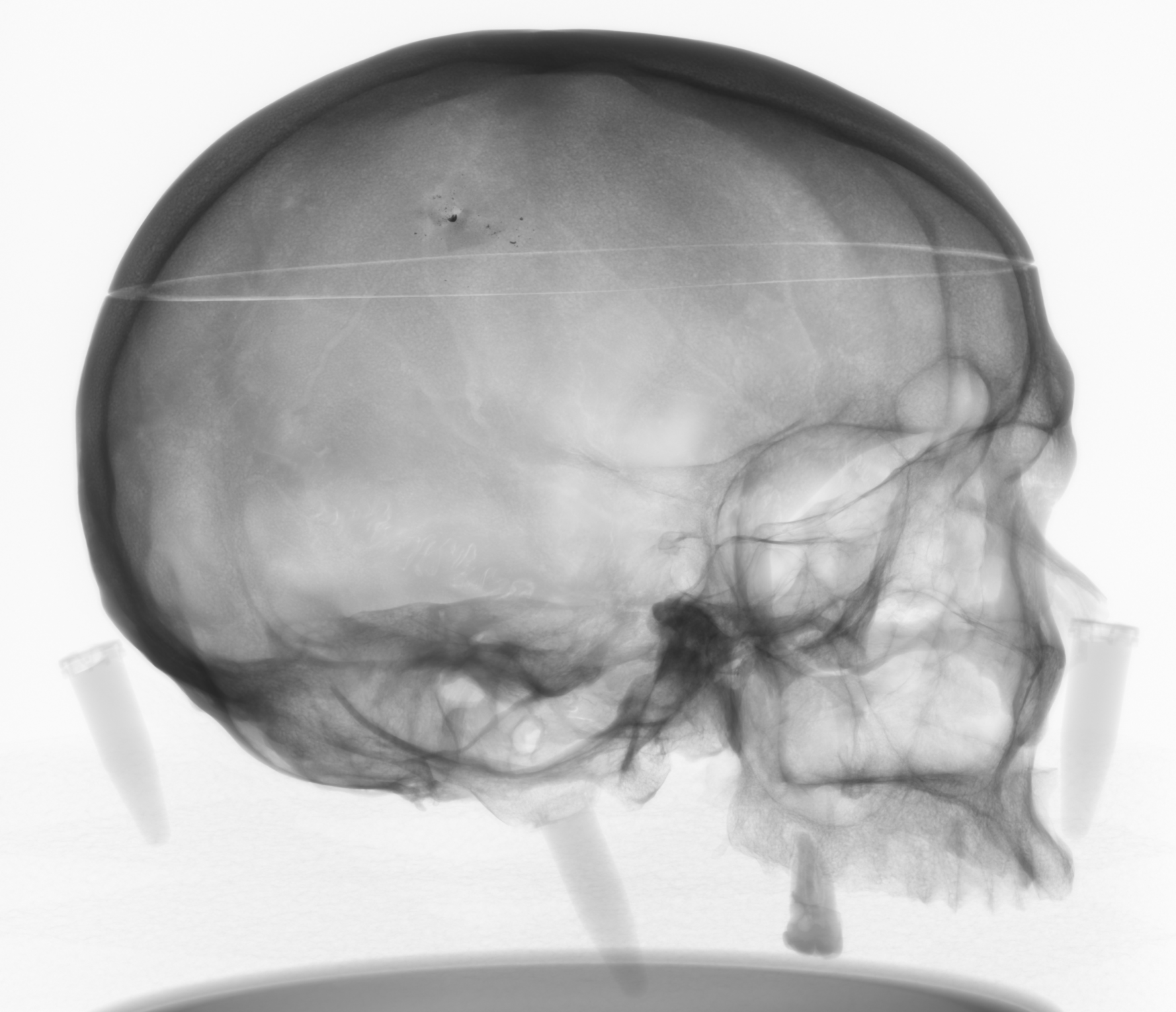}
    \caption{Raw data example -- one of the 2501 2D cone beam X-ray projection images.}
\end{figure}

\paragraph{Raw data}
\texttt{`halle\_skull\_2D\_projections.zip`} contains 2501 projection TIFF images, recorded by a cone beam micro-CT as described in section 3.

\paragraph{Reconstructed 3D volume}
\texttt{`halle\_skull.nrrd`}
is a nearly raw raster data (NRRD) \cite{NRRD} file containing the reconstructed 3D micro-CT. The file was produced by 

\begin{enumerate}
    \item reconstructing a volume with the proprietary software CT Pro (Metris, Tring, UK) using the settings defined in the file \texttt{`etc/halle\_skull\_3D\_reconstruction.xtekVolume`}, then
    \item converting the CT Pro output to NRRD using ImageJ \cite{ImageJ}, and then
    \item scaling the reconstruction to approximate HU.
\end{enumerate}
HU in a voxel are defined by $\mathrm{HU} = 1000 \times (\mu_\mathrm{voxel}-\mu_\mathrm{water}) / (\mu_\mathrm{water}-\mu_\mathrm{air})$. The absorption values $\mu_\mathrm{air}$ for air and $\mu_\mathrm{water}$ for water were manually segmented from three embedded water samples and surrounding air.
Note that these HU are only approximated -- as this is an ad hoc calibrated cone beam CT, the resulting HU values should not be interpreted as quantitative \cite{Mah}.

The finished 3D volume has a isotropic spacing of \SI{0.125}{\milli\meter} with 1150 $\times$ 1700 $\times$ 1400 16-bit signed integer voxels.

\paragraph{Bone segmentation}
\texttt{`halle\_skull\_segmented.nrrd`} is a semi-manual segmentation of the skull bone based on \texttt{`etc/threshold.py`} which combines a fixed threshold and a Sauvola binarization \cite{Sauvola}. Incorrectly segmented voxels, such as reconstruction artifacts, were cleaned up from the segmentation by hand using MITK \cite{MITK}.

\section{Subject}
The subject is an \emph{ex vivo} skull of a body donor, a 70 year old male. He signed a declaration of consent at lifetime for the use of his body for scientific purposes. The skull was provided by the Institute of Anatomy and Cell Biology, Martin-Luther-University Halle-Wittenberg, Halle (Saale), Germany. Other than the calcifications around a small trepanation in the right parietal bone, this is a typical human skull. There is one coronal cut, which was necessary during preparation of the skull. The skull was imaged without jaw bone.

\section{Measurement}
X-ray tomography was performed with an industrial cone beam micro-CT (XT H 225, Nikon Metrology) with \SI{150}{\kilo\volt}, \SI{400}{\micro\ampere} and using a \SI{1}{\milli\meter} copper filter. The detector had 1750 $\times$ 2000 pixels with \SI{200}{\micro\meter} isotropic spacing. This and other metadata of the acquisition can be found in the files \texttt{`etc/halle\_skull.xtekct`} and \texttt{`etc/halle\_skull.ctprofile.xml`}.
2501 angles were measured, each by averaging 8 exposures with \SI{708}{\milli\second} exposure times. This covered a \SI{360.349}{\degree} rotation -- exact angles are listed in the file \texttt{`halle\_skull\_2D\_projections/\_ctdata.txt`}.

The skull was placed on a low-density polyethylene packing foam together with three water filled \SI{1.5}{\milli\liter} polypropylene micro-centrifuge tubes. The cut-off top of the skull was held in position by \SI{48}{\micro\meter} thick polypropylene tape.

\section*{Acknowledgments}
Thanks to Prof.~Heike Kielstein of the Institute of Anatomy and Cell Biology of the Martin-Luther-University Halle-Wittenberg, Halle (Saale), Germany, for providing the sample.
Thanks to Dr.~John Maximilian Köhne of the Department of Soil Physics, Helmholtz-Zentrum für Umweltforschung UFZ, Halle (Saale), Germany, for his essential contributions in the micro-CT data acquisition.
Thanks to Prof.~Hans-Jörg Vogel of the Department of Soil Physics, Helmholtz-Zentrum für Umweltforschung UFZ, Halle (Saale), Germany, for providing access to the micro-CT of his department. Thanks to Prof.~Jan Laufer of the Fachgruppe Medizinische Physik, Institut für Physik, Martin-Luther-Universität Halle-Wittenberg, Halle (Saale), Germany, for hosting this research.

This work was funded by the German Research Foundation (DFG, Deutsche Forschungsgemeinschaft) under grant number 471755457.

\end{document}